\documentclass[twocolumn,twoside,slac]{revtex4}
\usepackage{graphicx}
\usepackage{fancyhdr}
\usepackage{array}
\usepackage[latin1]{inputenc}
\newcommand{\argminbetawiggle}
  {\mathop{\text{argmin}}_{\vec{x}}}
  \newcommand{\betawiggle} {\vec{x}} 
  
  \newcommand{\betawiggleahat }[1] {\hat{\vec{x}}_{#1}}

\def\nima#1#2#3  {{Nucl. Instr. and Meth. A} {\bf#1} (#2) #3.}

\begin{document}

\title{New vertex reconstruction algorithms for CMS}

%

%
\author{R. Fr\"uhwirth, W. Waltenberger}
\thanks{Supported by the Fonds zur
F\"orderung der wissen\-schaft\-lichen Forschung, Project 15177. Corresponding author.}
\affiliation{Institut f\"ur Hochenergiephysik der \"OAW, Vienna, Austria}
\author{K. Prokofiev, T. Speer}
\affiliation{Physik-Institut der Universit\"at Z\"urich, Switzerland}
\author{P. Vanlaer}
\thanks{Supported by the Belgian
Federal Office for Scientific, Technical and Cultural affairs
through the Interuniversity Attraction Pole P5/27.}
\affiliation{Interuniversity Institute for High Energies, ULB, Belgium}
\author{E. Chabanat, N. Estre}
\affiliation{Institut de Physique Nucléaire de Lyon, Villeurbanne, France}

\begin{abstract}

The reconstruction of interaction vertices can be decomposed into a pattern
recognition problem (``vertex finding'') and a statistical problem (``vertex
fitting''). We briefly review classical methods.  We introduce novel
approaches and motivate them in the framework of high-luminosity experiments
like at the LHC.  We then show comparisons with the classical methods in
relevant physics channels.

\end{abstract}

\maketitle

\thispagestyle{fancy}


\section{INTRODUCTION}

Vertex reconstruction algorithms face new challenges in high-luminosity
scenarios such as the LHC experiments. Vertex finding algorithms have to be able
to disentangle the tracks of vertices in difficult topologies, such as from
decay vertices which are very close to the primary vertex or decay chains with
very small separations between the vertices. Vertex fitters will need to be
robustified, since outliers and non-Gaussian tails in the distributions of the
errors of the track parameter will occur frequently.

We pursue extensive studies of vertex reconstruction algorithms that are
capable of dealing with ambiguities and track mis-reconstructions.
Section~\ref{sec_fitting} discusses robustifications of vertex fitting algorithms.
Section~\ref{sec_finding} presents novel approaches to the vertex finding
problem, derived from the clustering literature.

\section{VERTEX FITTING}
\label{sec_fitting}
Robustified vertex fitting has already been discussed in \cite{moscow}; we shall
only briefly review this topic here.

The classical methods in this field are {\it least-square methods}. The {\it
breakdown point} of LS estimators is zero, which means that even a single outlier
track can bias the resulting fit significantly. For noisy environments such as
the LHC experiments robustifications of the classic LS methods were
investigated. We suggest three new methods:
\begin{itemize}
 \item{}Adaptive method:
        Instead of minimizing the sum of residuals we
        minimize a weighted sum of squared residuals:
        $$ \betawiggleahat{\text{Adaptive}} =
		   \argminbetawiggle \sum_{i=1}^{n} \left( w_i \cdot r_i^2 ( \betawiggle ) \right) $$
		Outliers are not discarded but downweighted
		according to the weight function
		\begin{equation}
        w_i = \frac{1}{1+ e^{ \left( r_i^2 - r_{\text c}^2 \right) \beta }}
		\label{eq_weights}
		\end{equation}
		Here $r_c$ denotes a cutoff parameter, while $\beta \equiv 1 / (2 T)$
		introduces a temperature that is reduced in each iteration
		step in a well-defined annealing schedule.
		An iterative weighted LS procedure is used to find this minimum.
 \item{}Trimming method:
		We minimize only a user-defined fraction of the sum of the squared
		residuals: $$ \betawiggleahat{\text{Trimming}} = \argminbetawiggle
		\sum_{i=1}^{h < n} r_i^2 ( \betawiggle )$$
		A fast method that finds this minimum is described in \cite{fast-lts}.
 \item{}LMS:
        We minimize the median of squared residuals:
		$$ \betawiggleahat{\text{LMS}} = \argminbetawiggle \text{med} \left( r_i^2 (
		\betawiggle ) \right)$$
		Only a simplified algorithm has so far been found that is compatible
		with our CPU constraints. This algorithm works separately on each
		coordinate of the points of closest approach of the tracks with respect
		to a vertex candidate.  This ignores the spatial structure of the data.
		A full 3d method that works within our CPU requirements is still
		searched for.
\end{itemize}

The conclusions that we draw are as follows \cite{jorgen}:
\begin{itemize}
 \item{}the adaptive method should be considered a good default method;
        it deals with a great many different situations in an optimal or nearly
        optimal way. It leaves clean vertices almost unaffected, while at same time
		it is a very robust algorithm.
 \item{} The trimming vertex fitter may be interesting if the number of
         outliers is known in advance. In any other situation it is inferior to the
		 adaptive method.
 \item{} The LS fit is the fastest 3D fit, and optimal for extremely pure data.
 \item{} The coordinate-wise LMS fit is the fastest method but it is very
         unprecise (a few hundred microns compared to a few tens
         for primary vertices). It can nevertheless be used to provide
         a first guess of the vertex position.
\end{itemize}

\section{VERTEX FINDING}
\label{sec_finding}
We categorise the set of vertex finding algorithms into hierarchic and
non-hierarchic methods. Hierarchic methods are algorithms whose workings can
be visualised with a dendrogram.  Hierarchic methods can further be split
into divisive and agglomerative methods.

Divisive methods start with one cluster that contains all tracks; after each
iteration a certain subset of tracks is split off from the cluster into its own
cluster, which may in turn itself be split into sub-clusters. All algorithms
stop until a certain formal criterion is met.

Agglomerative methods start assigning a singleton cluster to every single
track. The most compatible clusters are then merged in every iteration step.
Again the procedure is stopped when a formal condition is met. The most decisive
factor in these methods is the {\it metric} that is employed to compute the
compatibility between two clusters.

Let $\alpha$ and $\beta$ denote two clusters.
Let further $s$ be the set of all minimum distances between
track pairs with one track in cluster $\alpha$ and the other in cluster $\beta$. We can now
choose as the metric e.g.:
\begin{equation}
  d(\alpha,\beta) = \text{min}(s), \text{max}(s), \bar{s}, \text{median}(s), \dots
\end{equation}

The choice $d(\alpha,\beta) = \text{min}(s)$ implements a {\it single linkage} or {\it minimum
spanning tree} procedure,
whereas $d(\alpha,\beta) = \text{max}(s)$ is often referred to as a {\it complete linkage}.

The following theorem significantly reduces the number of reasonable choices:

\begin{figure}[h!t]
\begin{center}
\includegraphics[width=150pt]{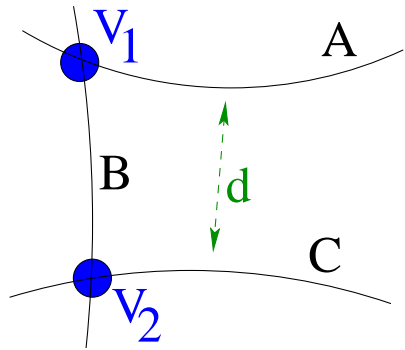}
\caption{Schematic description of how the triangle inequality is violated in
the track clustering problem.} \label{fig_tri}
\end{center}
\end{figure}
{\bf Theorem:} The triangle inequality does not generally hold for the minimum
distances between a set of $n$ tracks.

{\bf Proof:} Let $A, B, C$ denote three tracks. Let $A$ and $B$ share one common
vertex $V_1$; let further $B$ and $C$ also share one common vertex $V_2$.
Then:
\begin{eqnarray}
 \overline{AB} = \epsilon, \overline{BC} = \epsilon,
 \overline{AC}=d \gg \epsilon\nonumber\\
 \rightarrow
 \overline{AB} + \overline{BC} \ll \overline{AC} \hspace{2mm}q.e.d.
 \label{eq_triangle}
 \end{eqnarray}

This means that the choice $d(\alpha,\beta) = \text{min}(s)$ would cluster
A, B and C into a single vertex. We can therefore safely discard single linkage 
from the list of promising algorithms.

Until now the best results were obtained with the choice $d(\alpha,\beta) =
\text{max}(s)$, i.e.  with a {\it complete linkage} procedure.

An alternative to the above metrices is of course to fit vertices for each
cluster with more than one track, and use these vertices as ``representatives''
of the cluster.

\subsection{Finding-Thru-Fitting}
\label{sec_pvr}
The most mature algorithm in CMS is the ``principal vertex reconstructor'',
also known as the ``finding-thru-fitting'' method.  It is a divisive method
that internally uses a fitter and a track-to-vertex compatibility estimator to
decide which tracks are to be discarded at each iteration step. The
maturity of the implementation and the algorithmic simplicity make it an ideal
baseline for performance evaluation.

\subsection{Apex points}
In order to overcome the topological problems described in section
\ref{eq_triangle}, we conceived another approach: the apex point formalism.
The main concept is that the tracks are substituted by representative points,
the {\it apex points}.  These points should fully represent the tracks with
respect to the vertex finding problem at hand.  The space that the apex points
are defined in can have any dimension; it must only be equipped with a proper
metric fulfilling the triangle inequality.  Our current implementation produces
three-dimensional points in a Euclidean space, together with a 3x3 error
matrix. Note that the apex-point-to-track mapping needs not be unique; it may
very well be necessary that $n$ apex points, $n > 1$, represent one track.

One can of course also formulate hierarchic clustering methods on top of the
apex points.

\subsection{Apex point finders}
An algorithm that searches for such representative ``apex'' points is called an
apex point finder.  Since these finders operate on the points of closest
approach, they can be formulated as generic pattern recognition problems in one
dimension (i.e.~along the tracks). Thus the set of potential apex point finding
algorithms is huge; a systematic effort to choose algorithms that satisfy our
needs is an ongoing process. So far we have only investigated a few simple
methods:
\begin{itemize}
 \item{}The HSM (half sample mode) finder iteratively calls an LMS estimator on the
 points of closest approach.
 \item{}The MTV (minimal two values) finder looks for the two adjacent points
 of closest approach whose sum of distances to their counterparts on the other
 tracks is minimal.
 \item{}The MAMF (minimum area mode) finder looks for the two adjacent points
 whose sum of distances to their counterparts times their distance is minimal.
\end{itemize}

Future research will try more sophisticated algorithms such as a deterministic
annealing \cite{anneal}, \cite{oldanneal} method, or a gravitational
clusterer \cite{gravity}.

\begin{figure}[h!t]
\begin{center}
\includegraphics[width=250pt]{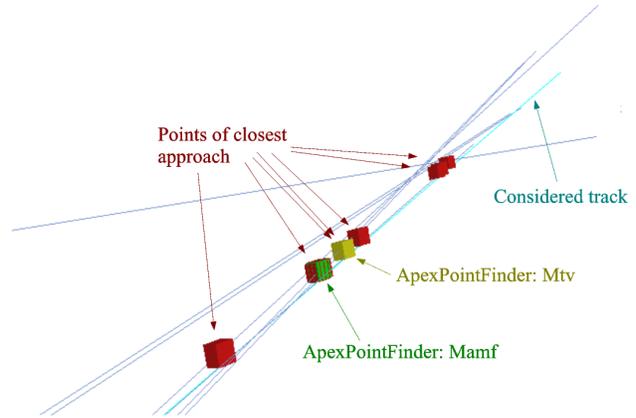}
\caption{Two apex point finders at work.}
\label{fig_apex}
\end{center}
\end{figure}

\subsection{Global association criterion}
The weights that have been introduced for the adaptive fitting method (see
section~\ref{eq_weights}) can also be used to define a global ``plausibility''
criterion of the result of a vertex reconstructor.  With $m$ being the total
number of tracks and $n$ being the number of vertices we define the global
association criterion (GAC) by:
$$
  p = \frac{1}{n\cdot m} \sum_{i=1}^{\text{m}} \sum_{j=1}^{\text{n}} p_{ij}
$$

where
$$
  p_{ij} = \left\{
	       \begin{tabular}{lc}
		   $1 - w_{ij}$ & if $i \in j$ \\
		   $w_{ij}$ & otherwise
		   \end{tabular}
		   \right.
$$
and $w_{ij}$ is the weight $w_i$ (\ref{eq_weights}) of track $i$ with respect
to vertex $j$.

The most important open question with respect to this criterion is how
it relates to the ``Minimum Message Length'' \cite{mel}. Can the
information theoretic limit of the vertex finding task be formulated in terms
of the GAC?

The potential uses of such a criterion are manifold:
\begin{itemize}
 \item{}Exhaustive vertex finding algorithm. All combinations of track clusters
 could at least in principle be iterated through, then one can decide for the
 smallest GAC found.
 \item{}Stopping condition. The GAC could also serve as a
 stopping condition in a wide range of algorithms.
 \item{}Super finder algorithms. One could also use it to resolve ambiguities.
 More than one vertex reconstructors could be used on one event, the GAC could
 then decide for the ``better'' solution.
\end{itemize}

Clearly, some more research in this direction will have to be done.

\subsection{Learning algorithms}
``Learning algorithms'' can easily be formulated on top of the apex points;
some can also work on the distance matrix itself. Good candidates for such
algorithms are:
\begin{itemize}
 \item{}Vector quantisation or the {\it k-means} algorithm, which have dynamic vertex
 candidates (``prototypes'') that are attracted by the apex points.
 \item{}Potts neurons \cite{potts} or the super-paramagnetic clustering
 algorithm \cite{spc}; these algorithms attribute a spin-state or a
 mathematical equivalent to every apex point. Spin-spin correlations together
 with an annealing schedule will then make sure that similar apex points are
 described by the same spin vector.
 \item{}Deterministic annealing \cite{anneal}; this method formulates the
 clustering problem as a thermodynamic system with phase transitions, each
 transition introducing a new separate cluster of apex points.
\end{itemize}

\subsection{Simulation experiment}
We compared one of our algorithms with two standard vertex finding procedures:
the PVR (see section~\ref{sec_pvr}) and the so-called D0Phi method \cite{d0phi},
\cite{cdf}
-- a special purpose algorithm based on the impact parameters of the tracks at
the beamline.  As a novel method to compare with we chose an agglomerative
clusterer with a vertex fits as ``representative points'', as it was explained
in the last paragraph of section~\ref{sec_finding}.
Our testing was done with 1000 Monte Carlo 50 GeV $b\bar{b}$ events, generated
with the CMSIM simulation program~\cite{cmsim}.  Before the actual comparison
all algorithms were automatically fine-tuned to maximize the following ``score
function'':
\begin{eqnarray}
  \text{Score} \equiv 10 \cdot \text{Eff}_{\text{Prim}} \cdot
  \text{Eff}_{\text{Sec}} \cdot \text{Pur}_{\text{Prim}}^{0.25} \cdot
  \text{Pur}_{\text{Sec}}^{0.25} \cdot \nonumber \\
  \text{AssEff}_{\text{Prim}}^{0.25} \cdot \text{AssEff}_{\text{Sec}}^{0.25}
  \cdot \left(1-\text{Fake}\right)^{0.5} \nonumber
\end{eqnarray}

``Eff'', ``Pur'', ``AssEff'', and ``Fake'' denote the performance estimators described in
\cite{pascal-chep2003}; ``Prim'' denotes primary vertices, ``Sec'' stands for
secondardy vertices.

\subsection{Results}
\begin{figure}[h!t]
\includegraphics[width=250pt]{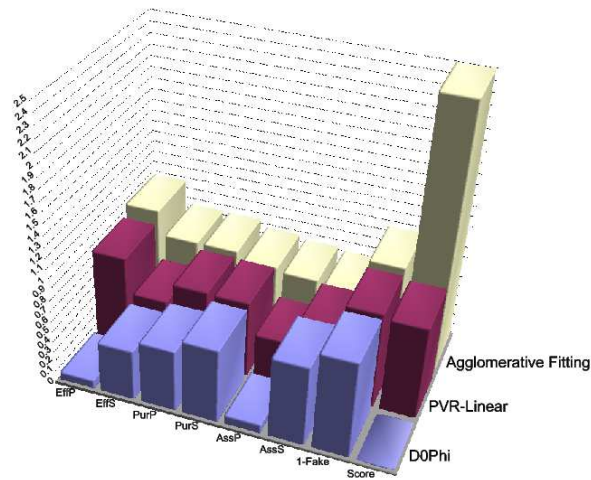}
\caption{Analysis of performance: one novel agglomerative finder compared to
older vertex finding algorithms. The agglomerative method has a much higher
secondary vertex finding efficiency, while it reports about the same fake
rate.}
\label{fig_tourn}
\end{figure}

In the inclusive secondary vertex finding scenario, the agglomerative fitting
procedure finds up to 80 percent of the secondary vertices, as opposed to the
50 -- 60 percent found by older algorithms. Note that the D0Phi algorithm is not
intended to find any primary vertices, hence the total score parameter is meaningless.
See figure~\ref{fig_tourn} for the complete comparison.

\section{CONCLUSIONS}
We have reached a good understanding of the robustification methods of the
classical LS vertex fitters. We suggest that the adaptive method be used as the
new default fitting procedure for CMS and possibly other experiments as well.
Surely, we still lack such an exhaustive understanding in the case of the much
more complex task of vertex finding, although here we were able to exclude certain
classes of algorithms on the basis of purely theoretic considerations.  Our
first results are most promising; we can quite clearly demonstrate that with
respect to e.g.~secondary vertex finding classical methods such as the
``finding-thru-fitting'' algorithm can be surpassed by far.  Future research
will mainly focus on three areas:

\begin{itemize}
 \item{}apex point finding algorithms,
 \item{}``learning'' algorithms,
 \item{}the global association criterion and its implications.
\end{itemize}


\end{document}